# A Critical Evaluation of the Role of the Precursor Complex and Counterions in Synthesis of Gold Nanoparticles in Micellar Media

Daniel Grasseschi,*[2] Filipe S. Lima,[1] Hernan Chaimovich,[1] Henrique E. Toma[1]

[1]Instituto de Química, Universidade de São Paulo. Zip Code 26077, CEP 05513-970, São Paulo, Brazil.

[2] Mackgraphe-Graphene and Nanomaterials Research Center, Mackenzie Presbyterian University, São Paulo, Brazil

*daniel.grasseschi@usp.br

## Abstract

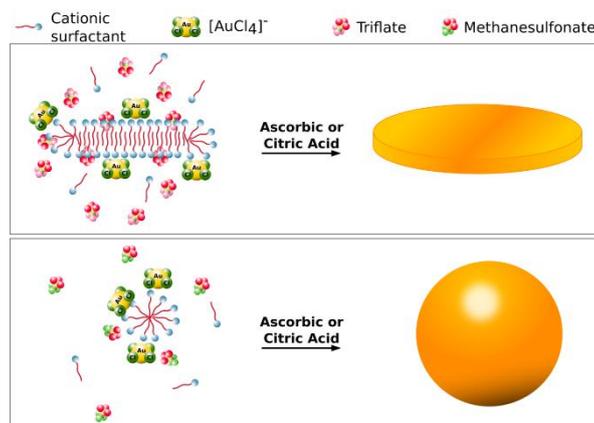

We demonstrate here that the gold nanoparticles (AuNPs) obtained in micellar solutions are susceptible to the precursor complex nature and specific counterion effects controlling the reaction kinetics, particles size and shape. The tested approach consisted in reducing the $HAuCl_4$ complex with ascorbic acid or citric acid in dodecyltrimethylammonium (DTA) solutions, at 318 K, in the presence of bromide (Br), methanesulfonate (Ms), trifluoroacetate (TFA) or trifluoromethanesulfonate (Tf) ions. Bromide ions replaced the chloride ions in the precursor gold complex, and no nanoparticle formation could be observed in DTA-Br media, in the absence of seeds. In contrast, scanning electron microscopy images and atomic force microscopy measurements showed that quasi-spherical AuNPs were formed in DTA-Ms and DTA-TFA solutions, while flat disk-like or triangular AuNPs were obtained in DTA-Tf solutions, with no need of adding seeds in all such cases. The results indicated that the reduction of the precursor gold complex takes place at the micellar interface and that the specific counterion effects should be taken into account for controlling nanoparticles formation in micellar media.

## Introduction

Gold nanoparticles (AuNPs) exhibit unique chemical and physical properties[1–6] associated with the surface plasmons oscillating in resonance with the electromagnetic waves, also referred as Localized Surface Plasmon Resonance (LSPR)[7]. Size, shape and chemical environment greatly influence LSPR, influencing directly the optical properties of the nanoparticles.[8–11] The design of metallic nanoparticles with different shapes has been pursued by many research groups in recent years,[12–14] and a special attention has been dedicated to nanoparticles with tips, such as rods and prisms, due to local enhancement of the electromagnetic field promoted by the lighting rod effect.[15] Such nanoparticles are enabling important applications in chemistry and biology as surface enhanced Raman scattering (SERS) substrates.[3]

Anisotropic AuNPs can be obtained by employing biological extracts as reducing and stabilizing agents.[16–19] However, in this case, the particles size and shape control is hampered by the matrix complexity and the usually long reaction times. The use of a polymeric matrix represents another important alternative,[20–22] although the use of a surfactant matrix is becoming more competitive for practical reasons, for instance, the possibility of using lower reaction temperatures and time.[23,24]

The synthesis of anisotropic gold nanoparticles (AuNP) in surfactant solutions has been performed by means of the seed-mediated method, using cetyltrimethylammonium bromide (CTA-Br) micelles as templates.[25,26] The rationale of this method is associated with the influence of the cationic anisotropic CTA-Br micelles, directing the growth of pre-synthesized isotropic nanoparticles, the seeds.[27] Murphy et al.[13] suggested that the anisotropic growth is governed by chemical and steric factors, such as the interaction between quaternary ammonium group and the {100} gold surface. This process is accompanied by the [AuBrCTA] complex formation. The shape control can be achieved by changing the gold/CATB molar ratio,[28] or by adding co-surfactants like hexadecylbenzyldimethylammonium bromide,[29] Ag[1+] [30] or I[1-] [14,31] ions. Recently Zeng et al showed that the shape control could be made controlling the interaction and the assembly into mesoscales structures of the CTAB micelles by changing the metal precursor nature and the CTAB concentration, extending the methodology to the synthesis of PdNP and PtNP besides the AuNP.[32]

It should be noticed that the micellar geometry depends on the nature of the surfactant's counterion.[33] Such specific ion effects have been reported a century ago by Hoffmeister, investigating the effect of different salts in the precipitation of egg proteins.[34] They have been reported in several systems,[35–37] in special self-associating colloids, such as micelles formed by charged surfactants, exhibit dramatic ion-specific changes.[33,38,39] At low surfactant concentration, dodecyltrimethylammonium (DTA) bromide or chloride form spherical micellar aggregates.[40,41] The increase in the surfactant concentration, however, leads to a sphere-to-rod transition in the micelles generated in DTA-Br media, but not in DTA-Cl. Sphere-to-rod transitions can be also induced by aromatic anions as counterions of cationic surfactants,[42] and the geometry of the aromatic anions can lead to very different effects.[39] Not only shapes and sizes, but several micellar properties, such as packing, degree of order and hydration at the micellar interface and hydrophobic core, can be affected by the nature of the counterions, as recently shown for micelles formed by DTA trifluoromethanesulfonate (triflate, Tf).[43–47]

DTA-Tf forms micelles with low charge, i.e., low degree of counterion dissociation ($\alpha$), high aggregation numbers ($N_{agg}$) and the geometry of the aggregates are discoidal,[44] an uncommon shape for cationic micelles. The addition of sodium triflate to DTA-Tf solutions induced macroscopic phase separation,[43] by lowering the inter-micellar repulsion due to the counterion binding.[47] However, curiously, anions with similar structure of Tf, such as methanesulfonate (Ms) and trifluoroacetate (TFA), formed spherical aggregates with DTA, instead of discs.[46]

Therefore, the nanoparticles formation in micellar media can also be strongly influenced by the nature of the counterions.[48,49,50,51] However, in addition to their influence on the micelles geometry and interfaces, there is another critical aspect for consideration, which is the chemistry of the precursor complex involved and its relation with the micellar media. This aspect has been systematically investigated in this work, by monitoring the spectral changes of the starting [AuCl$_4$]$^-$ complex and the small angle X-ray scattering behavior after the [AuCl$_4$]$^-$ addition to DTA micelles in the presence of bromide, methanesulfonate, trifluoroacetic and triflate counterions. As reported in this paper, a contrasting behavior has been observed for the bromide species, which is a strong ligand for Au(III) ions, in relation to the non-coordinating Ms, TFA and Tf ions, highlighting the important role of the specific counterion effect associated with the precursor complexes and the micelles properties. These results suggest a new approach for the design and production of gold nanoparticles.

## Experimental

**Materials.** Trifluoromethanesulfonic acid, dodecyltrimethylammonium bromide (DTA-Br), trifluoroacetic acid, methanesulfonic acid, NaOH, tetrachloroauric acid (HAuCl$_4$), ascorbic acid, and citric acid (Sigma-Aldrich, St Louis Mo) were analytical grade and used without further purification. Dodecyltrimethylammonium triflate (DTA-Tf) and NaTf were prepared as previously described.[23] Water was deionized and distilled.

**Methods. Spectrophotometry.** Aliquots of the HAuCl$_4$ stock solutions were added to DTA-Tf solutions and ascorbic or citric acid was added after 15 minutes. The reactions were followed in a Hewlett Packard 8453A diode-ray spectrophotometer, in the 190 to 1100 nm wavelength range, using a quartz cuvette. HAuCl$_4$ reduction was followed at 317 nm. Apparent rate constant of HAuCl$_4$ reduction was determined by a single exponential decay fit to the kinetic curves before the growth of AuNPs, to avoid the contribution of baseline increase.

**Scanning Electron Microscopy, SEM.** After removing the excess of surfactant by centrifugation at 1400 rpm for three minutes, the gold nanoparticles suspension was deposited on highly ordered pyrolytic graphite, and dried under vacuum. The images were obtained using a JEOL model 7200 field emission electron microscope.

**Small Angle X-ray Scattering, SAXS.** 0.1 M DTA-Tf in water and with 0.004 M HAuCl$_4$ were added into quartz capillaries (2 mm of inner diameter) and experiments were performed in a Nanostar (Brucker) with the X-ray tube operating at 40 kV and 30 mA, producing Cu K$\alpha$ radiation ($\lambda$ = 1.54 Å). The total acquisition time of each scattering curve was equal to 3 h, and data was collected by a two-dimensional detector (HiSTAR). The two-dimensional scattering pattern was integrated to generate a profile of intensity as a function of the scattering vector (q = 4$\pi$sin$\theta$/$\lambda$), ranging from 0.014 to 0.35 Å$^{-1}$, using 2$\theta$ as the scattering angle. The sample-detector distance was 679 mm and experiments were performed at 318 K. SAXS curves were fitted by a model of flat membrane,[52,53]

$$I(q) = \frac{c}{q^2} [\int_{-d/2}^{+d/2} \rho(x) \cos(qx) \, dx]^2 \quad (1)$$

where c is a constant, d is the thickness of the bilayer, $\rho(x)$ is the electron density perpendicular to the bilayer plane. This equation was solved with two levels of electron density, with one region defined as the hydrophobic core of the bilayer (thickness $R_{core}$ and electron density $\rho_{core}$) and another region related with the polar region of the micelle (thickness $R_{shell}$ and electron density $\rho_{shell}$).

**Gold nanoparticles synthesis.** Regardless the concentration of reactants and surfactants, the AuNPs were synthesized with the following procedure: aliquots of HAuCl$_4$ were added to DTA-X aqueous solution at (318 K) with gentle mixing. The solution was maintained at 318 K for 15 minutes, and an aliquot of ascorbic or citric acid was added to the gold/surfactant solution, leading to the gradual development of the purple color associate with the AuNPs.

## Results and Discussion

### *General behavior of the gold nanoparticles in DTA-X micellar media.*

The electronic spectra shown in Figure 1 were recorded after adding $HAuCl_4$ and ascorbic acid (AA) to the DTA-X micellar solutions containing X = Br, Ms, TFA and Tf counterions, after 20 min of reaction, at the following (mmol L$^{-1}$) proportion: $HAuCl_4$(0.5)/AA(0.6) /DTA-X(50).

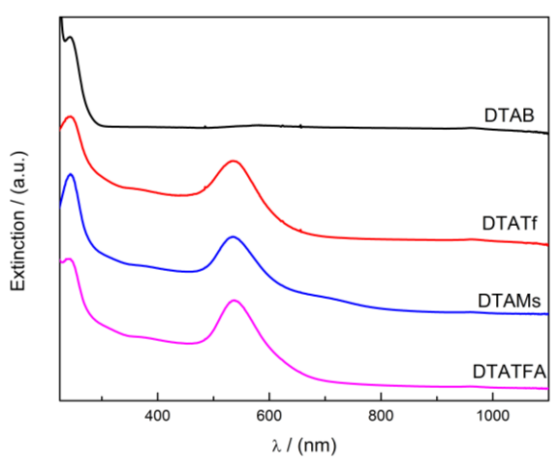

**Figure 1.** Extinction spectra of $HAuCl_4$ solutions in (black) DTAB; (red) DTATf; (blue) DTAMs; and (magenta) DTATFA after 20 minutes of AA addition.

The first striking result in Figure 1 is the occurrence of characteristic surface plasmon band at 535 nm in the DTA-Ms, DTA-TFA an DTA-Tf systems, indicative of the formation of gold nanoparticles, even without adding seeds, as normally required in this type of procedure. In contrast, no evidence of gold nanoparticles is observed in DTA-Br media.

The shapes and sizes of the generated AuNPs were also quite distinct, as shown in Figure 2. The nanoparticles synthesized in DTA-Tf were planar (disk-like or triangular), with average diameters of 26.9 nm and thickness of 2.6 nm. Additional experiments have shown that the diameter can be controlled between 20 and 70 nm and the thickness between 1 to 15 nm, depending of the gold concentration, as will be discussed later on the text. The observed geometry is consistent with the disk-like micelles observed in the DTA-Tf solutions,[44] indicating that the nanoparticles formation should take place at the interfacial regions. On the other hand, the AuNPs obtained with DTA-Ms and DTA-TFA were nearly-spherical, with average diameters of 32.2 and 43.3 nm, respectively. Thus, the surfactant´s counterion determines not only the feasibility of the AuNP synthesis in cationic surfactant solutions, but also the shapes and sizes of the resulting particles.

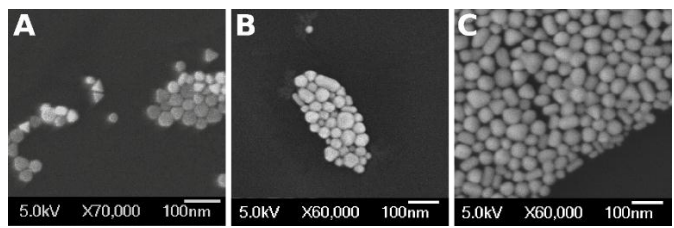

**Figure 2.** SEM images of AuNPs synthesized in: (A) DTA-Tf; (B) DTA-Ms; (C) DTA-Tf.

### *Ligand exchange in the [AuCl$_4$]$^-$ complex in NaX solutions*

In order to understand the distinct results observed in the DTA-Br system, the chemical behavior of the [AuCl$_4$]$^-$ complex has been investigated in the presence of the Ms, TFA an Tf counterions, as shown in Figure 3. In this case, the concentration of [AuCl$_4$] used was 2.5 x 10$^{-5}$ M to avoid saturation.

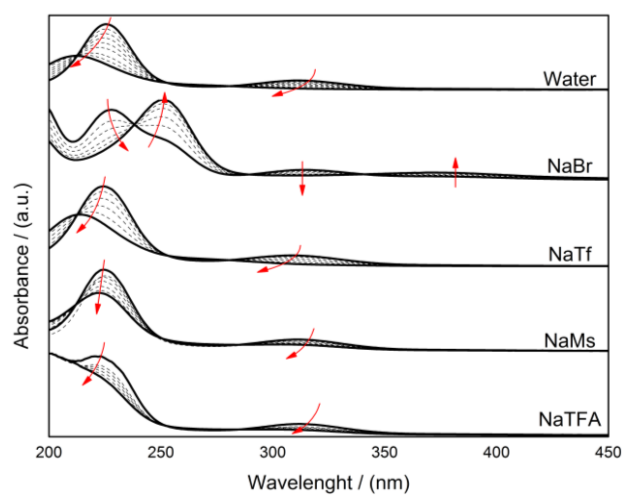

**Figure 3.** Time evolution of the electronic spectra of $HAuCl_4$ 2.5 x 10$^{-5}$ M in water, NaBr, NaTf, NaMs and NaTFA 0.05 M solutions. Arrows indicate the time evolution.

The [AuCl$_4$]$^-$ complex exhibits two characteristic bands at 220 and 320 nm, ascribed to ligand-to-metal charge-transfer bands, LMCT (Cl → Au),[54] as shown in Figure 3(top). In pure aqueous solution, such bands undergo a systematic decay, consistent with the replacement of the chloride ligands by the water molecules. Because of the high acidity of the $Au^{3+}$ ions, the coordinated water molecules should be deprotonated, yielding successive complexes of the type [AuCl$_x$(OH)$_{4-x}$]$^-$.[50,54,55] The spectra recorded in pure water and in NaTf, NaMS and NaTFA solution are very similar (Figure 3), confirming the weak coordinating properties of such counterions. In NaTFA solutions, the TFA absorption around 200 nm partially overlaps the gold complex transition at 225 nm, but the observed changes at 312 nm reproduced the same complex dissociation profile.

However, in NaBr/[AuCl$_4$]$^-$ solutions, the original charge-transfer bands are gradually converted into new strong bands at 251 and 374 nm, consistent with the exchange of chloride by bromide ions, yielding [AuCl$_{4-x}$Br$_x$]$^-$ complexes. The spectral changes are consistent with the stronger π-donor character of the bromide ions. As a consequence, the redox potential of the final [AuBr$_4$]$^-$ complex decreases, slowing down the reduction rates with ascorbic acid, thus hampering the formation of the gold nanoparticles. In contrast, in the presence of non-coordinating counterions (Ms, TFA, Tf) the aquation of the [AuCl$_4$]$^-$ complex yields more oxidizing precursors, facilitating their reduction even with mild oxidizing agents, and low temperatures. The presence of a hard base such as Tf, Ms and TFA, which does not interact with the gold complex, was found to eliminate the need of seeds. Therefore, the ligand exchange at the coordination shell of the gold precursor complex should play an important role in the AuNPs formation.

*Ligand exchange in the [AuCl$_4$]$^-$ complex in DTA-X micellar media.*

In order to evaluate the influence of the micelles in the chemistry of the precursor complexes, their ligand substitution behavior were reproduced in the presence of the DTA-X systems, as summarized in Figure 4. However, in this case the concentration used was the same as the used in the nanoparticles synthesis and in this conditions only the LMCT band at 320 nm can be analyzed.

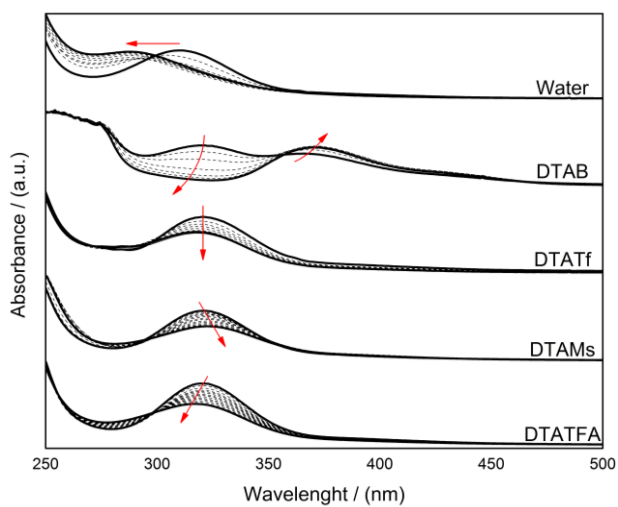

**Figure 4.** Time evolution of the electronic spectra of [AuCl$_4$]$^-$ in water, DTA-Br, DTA-Tf, DTA-Ms and DTA-Tf 0.05 M solutions. Arrows indicate the time evolution.

As shown in Figure 4, the [AuCl$_4$]$^-$ complex aqueous solution, undergoes partial dissociation of the chloride ligands, shifting the 320 nm LMCT band to 270 nm, corresponding to formation of [AuCl$_x$(OH)$_{4-x}$]$^-$ species. However, in the DTA-Tf solution, only an intensity change is observed at 320 nm (Figure 4). Such spectral behavior is rather informative, since it means that the replacement of the chloride ions by water is not occurring, in comparison with Figure 3. Instead, the lack of chloride/water exchange indicates that the [AuCl$_4$]$^-$ complex is located in a region of a poor water content, such as at the micellar interface. This is quite plausible, since the DTA-Tf interface is known to be considerably dehydrated.[44,45,46] A similar behavior is also reproduced in the DTA-Ms and DTA-TFA systems, with minor differences associated with their slightly more hydrated micellar interfaces.[46]

In the HAuCl$_4$-DTA-Br system (Figure 4), because of the presence of bromide ions, the observed spectroscopic changes are consistent with chloride/bromide ligand exchange, generating [AuCl$_{4-x}$Br$_x$]$^-$ species at the micellar interface. This interpretation is consistent with the slight shifts of the [AuBr$_4$]$^-$ bands, in relation to Figure 3, reflecting a different chemical environment promoted by the micelles.

*Gold complex reduction in the DTA-X micellar media*

In the absence of the micelles, the aquation of the [AuCl$_4$]$^-$ complex leads to [AuCl$_{4-x}$(H$_2$O)$_x$]$^n$ species in acid-base equilibrium with their corresponding hydroxo-species. The complexes undergo a fast reduction with ascorbic acid, generating the gold nanoparticles; however, plasmon band is broad indicating a poor size distribution, Figure 5. In the absence of stabilizing agents, aggregation takes place leading to a strong plasmon coupling band around 900 nm, followed by particle precipitation (Figure 5).

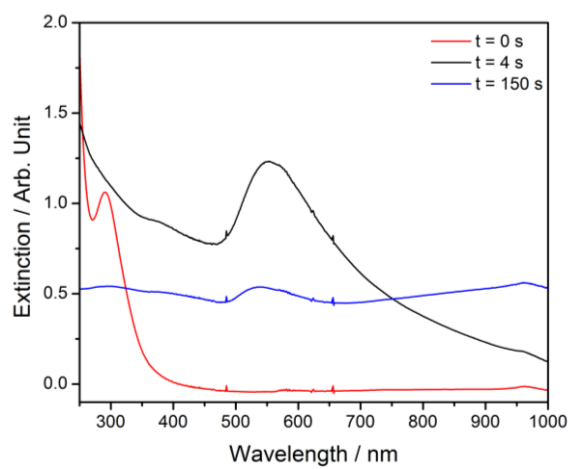

**Figure 5:** Extinction spectra of HAuCl$_4$ solution after 150 seconds of AA addition.

In the presence of the DTA-X micelles, the general behavior for the formation of the gold nanoparticles has already been shown in Figure 1. The comparison of the time evolution behavior in Figure 6, indicates that the formation of the gold nanoparticles proceeds very rapidly in the case of the DTA-Ms, DTA-TFA and DTA-Tf systems, while no reaction if observed in DTA-Br. This behavior suggest a burst nucleation in one single step flowed by one growth process, which should be controlled by the gold diffusion. In Figure 6, the behavior in pure aqueous solution reflects the formation and aggregation of the gold nanoparticles, and the comparison highlights the stabilizing effect promoted by the micellar media.

The gold complex present a distinct behavior on micellar systems with bromide as counters ion. Most of the synthesis of gold nanoparticles are usually carried out in CTA-Br (also denoted CTAB), and require the use of nanoparticles seeds in order to catalyze the gold reduction reaction. As we have shown in this work, in the presence of bromide ions the $[AuCl_4]^-$ complex is converted into $[AuBr_4]^-$ species, displaying a lower reduction potential,[56] thus decreasing the rates of electron transfer and of formation of the gold nanoparticles, Figure 6. For this reason, the use of nanoparticle seeds is required in the CTA-Br synthesis. In contrast, in the case of the non-coordinating counterions, the $[AuCl_4]^-$ complex is preserved at the surfactant interfaces, and the formation of the gold nanoparticles can proceed directly, with no need of seeds, as we have demonstrated in this work. Bromide ions have also been suggested to interfere, by specific coordination at certain nanocrystal facets, slowing down the nanoparticles growth. For instance, Murphy and coworkers[13,57] have explained the observed crystal anisotropic growth by the interaction between the CTA quaternary ammonium group and the {100} gold surface, intermediated by the [Au-Br-CTA] complex.

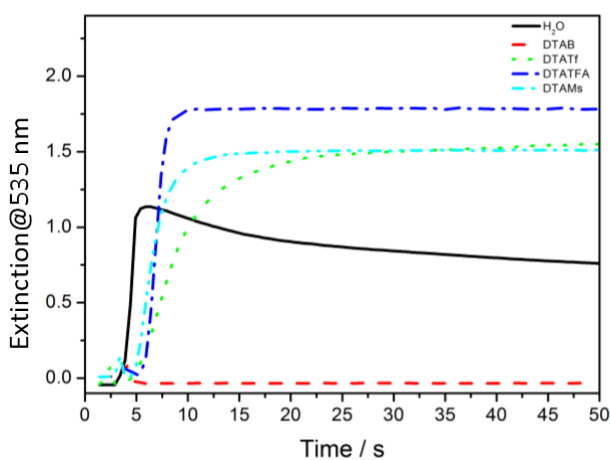

**Figure 6.** Extinction at 536 nm as a function of time for the five systems in (-) water; (--) DTAB, (···) DTATf, (-··) DTATFA and (-···-) DTAMs.

The time evolution for formation of gold nanoparticles in the DTA-Ms and DTA-TFA media is faster than that observed in DTA-Tf. This may be due to the presence of $[AuCl_{4-x}(H_2O)_x]^n$ ions in the relatively more hydrated micellar interface, which are more rapidly reduced than the $[AuCl_4]^-$ species. Planar particles is known to be thermodynamic unfavorable and are produce mainly by an kinetic control,[12,57] thus this slow growth rate in the DTA-Tf solution can favor the formation of planar particles, beside the micellar geometry. As a matter of fact, recently, Mirkin et al. discussed the growth mechanism of the AuNPs in seed-mediated methods, by focusing on the kinetic control and selective passivation of specific crystal facets.[12] Accordingly, the kinetics control can be achieved by changing the reduction rates of the gold complexes. Thus, the use of Br, Tf, Ms and TFA as DTA counter ion is a good example of kinetic control of the nanoparticle growth by controlling the species at the gold coordination sphere.

The precursor complexes should be distributed between the bulk and the micellar interface, according to their equilibrium constants[38], depending upon the structure of the species. Large ions[58] and organic molecules[59] are prone to preferentially binding to the micellar interface. Due to the relative small surface area of the aggregates, local concentration of molecules at the micellar interface can be considerably greater than those at bulk solution. Although the total fraction of molecules adsorbed at interface increases with the surfactant concentration, for a constant amount of precursor complexes, their average distribution at the micellar interface should decrease.

In order to evaluate this aspect, the reduction of the $[AuCl_4]^-$ complex at the micellar interface, was monitored at 317 nm as a function of the DTA-Tf concentration. The kinetics involved in the presence of ascorbic acid was too fast for monitoring using conventional techniques, but the few collected data at high [DTA-Tf] concentrations were in accord to our expectations, as shown in Figure 7A. In order to provide more reliable results, ascorbic acid was replaced by a more weakly reducing agent, such as the citrate ions. The kinetics, in this case, proceeds rather slowly and can be accurately monitored, showing a systematic decay of the $k_{app}$ with the DTA-Tf concentrations, as illustrated in Figure 7B. Therefore, the $[AuCl_4]^-$ reduction should be occuring at the micellar interface, since for the bulk reaction the rates would be independent of [DTA-Tf].

It should be notice that the slower reduction rate using citric acid as reducing agent favor the formation of bigger hexagonal and triangular planar disk-like particles, Figure 8. The $[AuCl_4]^-$ presence at the micellar interface and the lower reducing power of the citric acid provide a good kinetic control of nanoparticles growth. Which allied with the template effect leads to the formation of planar particles as main product.

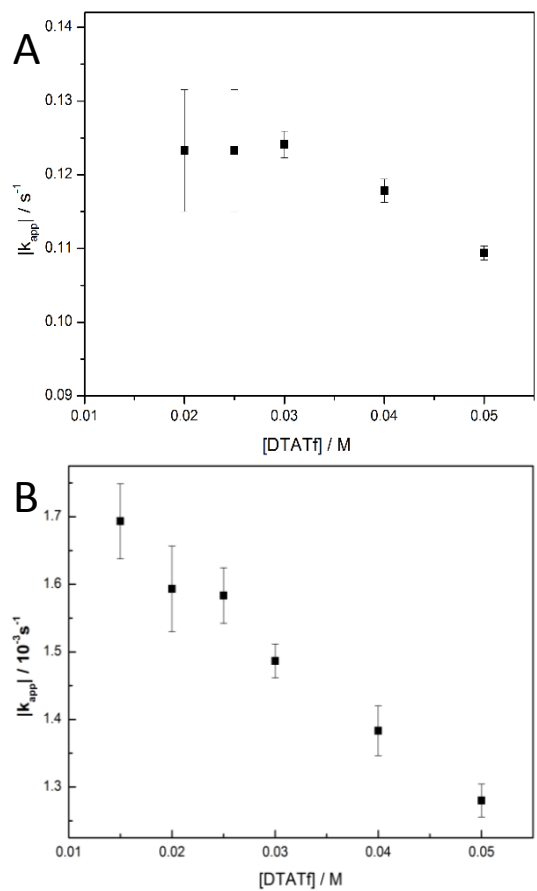

**Figure 7.** $k_{app}$ as a function of [DTA-Tf] using (A) ascorbic acid and (B) citric acid as reducing agent.

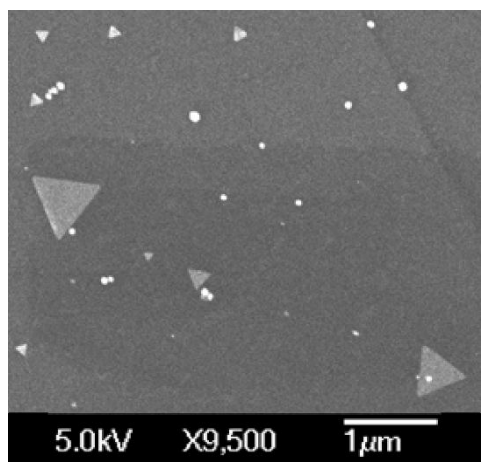

**Figure 8.** SEM images of AuNPs synthesized in: DTA-Tf using citric acid as reducing agent

### *The role of the counterions and complexes in the nanoparticles geometry*

The formation of AuNPs seems to be determined by the properties of the gold complex, while the shape of the gold particles seems to be related to the geometry of the micelles. As a matter of fact, the AuNPs synthesized in DTA-X exhibit the same geometry of the micellar aggregates formed by the same surfactants, e. g. disk-like planar structures for DTA-Tf and nearly-spherical structures for DTA-Ms and DTA-TFA.

Because salt-induced shape transitions of micellar aggregates are of common occurrence, especially with ions that possess high affinity to the micellar interface,[39,41,44,46] such as $[AuCl_4]^-$,[60] the addition of $HAuCl_4$ to the micellar solution could lead to shape changes of the micelles. For this reason, small angle x-ray scattering (SAXS) experiments were here performed in order to determine if the added $HAuCl_4$ complex can induce shape changes in the DTA-Tf aggregates.

The SAXS experiments illustrated in Figure 9 were focused on the DTA-Tf micelles in water, before and after adding $HAuCl_4$. A broad peak, centered at ~0.2 Å$^{-1}$, was observed in the SAXS curve of the two systems, as previously reported for the pure surfactant.[44] The position of this peak is coarsely related with the total thickness of the micellar aggregate, providing an estimative of the micellar thickness of ~$2\pi/(0.2$ Å$^{-1}) \sim 31$ Å. The SAXS curve of pure DTATf aggregates can be satisfactorily described by a two-level infinite bilayer model, which accounts for the thickness (but not the edge) of a disk.[44] The two-level infinite bilayer model is also reproduced the DTATf + $HAuCl_4$ SAXS curve, indicating that the addition of the gold complex did not induced shape changes in the DTATf aggregates. The parameters (Table 1), obtained by fixing $\rho_{water} = 0.33$ e/Å$^3$, are similar to those previously reported for this model.[44] Thus, the structure of the DTATf aggregates remained unchanged after the addition of the gold complex.

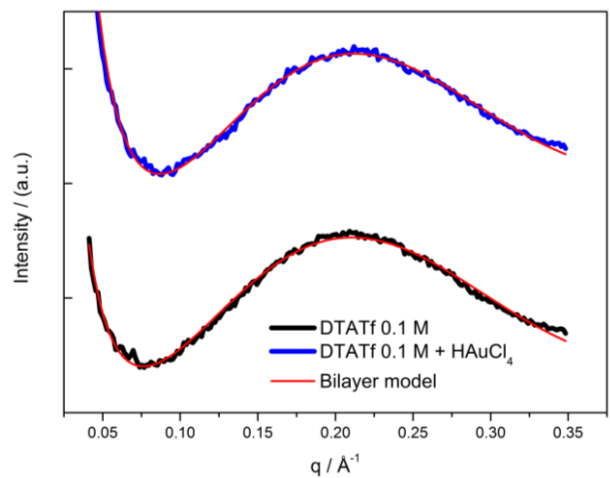

**Figure 9.** SAXS curve of 0.1 M DTATf in water and with added 0.004 M $HAuCl_4$.

**Table 1.** Parameters from Fit of the Two-Step Infinite Bilayer Model to the 0.1 M DTATf in Water and with Added 0.004 $HAuCl_4$ or $HPtCl_4$

| System | $R_{shell}$ / Å | $R_{core}$ / Å | $\rho_{shell}$ / e/Å$^3$ | $\rho_{core}$ / e/Å$^3$ |
|---|---|---|---|---|
| DTATf | 6 | 8 | 0.40 | 0.29 |
| DTATf + $HAuCl_4$ | 6 | 8 | 0.42 | 0.29 |

Another relevant question is whether the shape of the synthesized nanoparticles can be influenced by the existing counterions in the micellar system, besides the template effect exerted by the micelles surfaces. As we have recently reported, using hyperspectral dark field microscopy,[61] the nanoparticles shape has a strong influence on the plasmon resonance spectra. Therefore, by monitoring the spectra during the growth of the nanoparticles, one can access the changes of geometry along the process. This can be seen in Figure 9.

When the AuNP growth mechanism in DTA-X micelles occurs according to a single or uniform step, one can expect a continuous increase of the plasmon resonance band, as observed for the DTA-TFA system in Figure 10B. In this case, only a well behaved extinction band occurs at 540 nm, increasing with the time, during the synthesis. However, in the case of the DTA-Tf system, an initial plasmon resonance band is observed at 560 nm, reaching an apparent maximum, before converting gradually into a symmetric peak at 540 nm (Figure 10A). According to our recent work on hyperspectral dark field microscopy, this is consistent with the initial formation of triangular disks, which is gradually converted into hexagonal ones. This can also be readily seen in Figure 2A.

The presence of sharp tips, in triangular particles, can explain the red shift of the plasmonic resonance due the large charge concentration at the tip.[61] For the DTA-TFA system, this shift is not observed (Figure 10B) reflecting a single step growth of nearly spherical AuNPs (Figure 2B and C). These two contrasting results should reflect the kinetic differences associated with the distinct counterions, as illustrated in Figure 6, controlling the rates, and preferential growth of the gold nanoparticles.

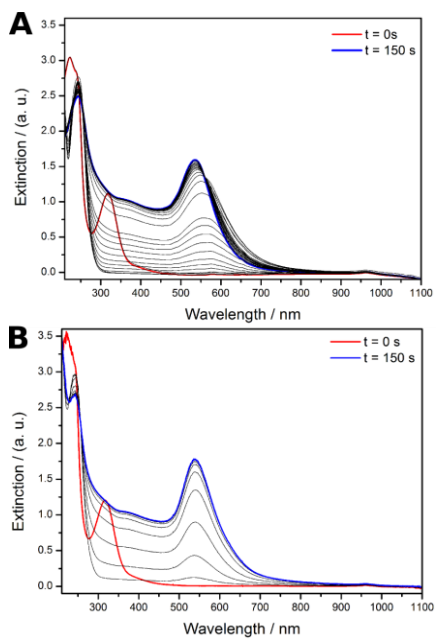

**Figure 10**: Extinction spectra of HAuCl$_4$ solutions in (A) DTA-Tf and (B) DTA-TFA after 150 seconds of AA addition.

*The effect of gold concentration in the nanoparticles geometry*

To evaluate the effect of gold concentration of the nanoparticles geometry the synthesis was made by the reduction of 0.25, 0.50, 0.75 and 1 mmol dm$^{-3}$ [AuCl$_4$]$^-$ with ascorbic acid in 50 mmol dm$^{-3}$ DTATf aqueous solution. The particles are here denoted AuND-25, AuND-50, AuND-75 and AuND-100, respectively.

As shown in Figure 11 A-D, the SEM images for AuND-X (X = 25, 50, 75 and 100) exhibit nearly round or hexagonally shaped nanoparticles, in addition to some triangular species. A particular microscopy image captured for AuND-100 (Figure 11E), exposed a fortuitous orientation of well defined arrays of nanodisks of about 15 nm thickness. The average diameter, measured over 500 nanoparticles, increased non-linearly from 25 to 53 nm for the AuND-25 to AuND-100 samples (Figure 11 F). The distribution width was narrower for AuND-25, AuND-50 and AuND-75 (exhibiting a standard deviation around 7 nm), in relation to AuND-100 (standard deviation = 14 nm) as shown curve D of Figure 11 F.

The thickness distribution of the gold nanodisks can be better seen in the AFM measurements, shown in Figure 12. The topographic AFM images indicate the formation of disk like geometries, exhibiting aspect ratios (diameter/thickness) varying from 40, for AuND-25, to 4 for AuND-100. The thickness distribution (Figure 12 E), based on phase contrast AFM images indicates only a small polydispersion.

The observed thickness of the AuNDs revealed that the particles size is determined by the dimensions and the packing properties of the micelle aggregates, as well as by the amounts of [AuCl$_4$]$^-$ employed in the preparations. At low [AuCl$_4$]$^-$ (0.25 mmol dm$^{-3}$), the thickness of the AuND particles (0.6 nm) was smaller than the total thickness of the DTATf aggregates, which is ca. 3 nm.[44] The increase in the gold complex concentration led to thicker aggregates, reaching 2.6 and 2.8 nm for the AuND-50 and AuND-75 nanoparticles, respectively, values which are close to the thickness of the disk-like DTATf aggregates.[46] However, a further increase in the [AuCl$_4$]$^-$ concentration, above 0.75 mmol dm$^{-3}$, led to a dramatic thickness increase, reaching 13.8 nm in the case of the AuND-100 nanoparticles.

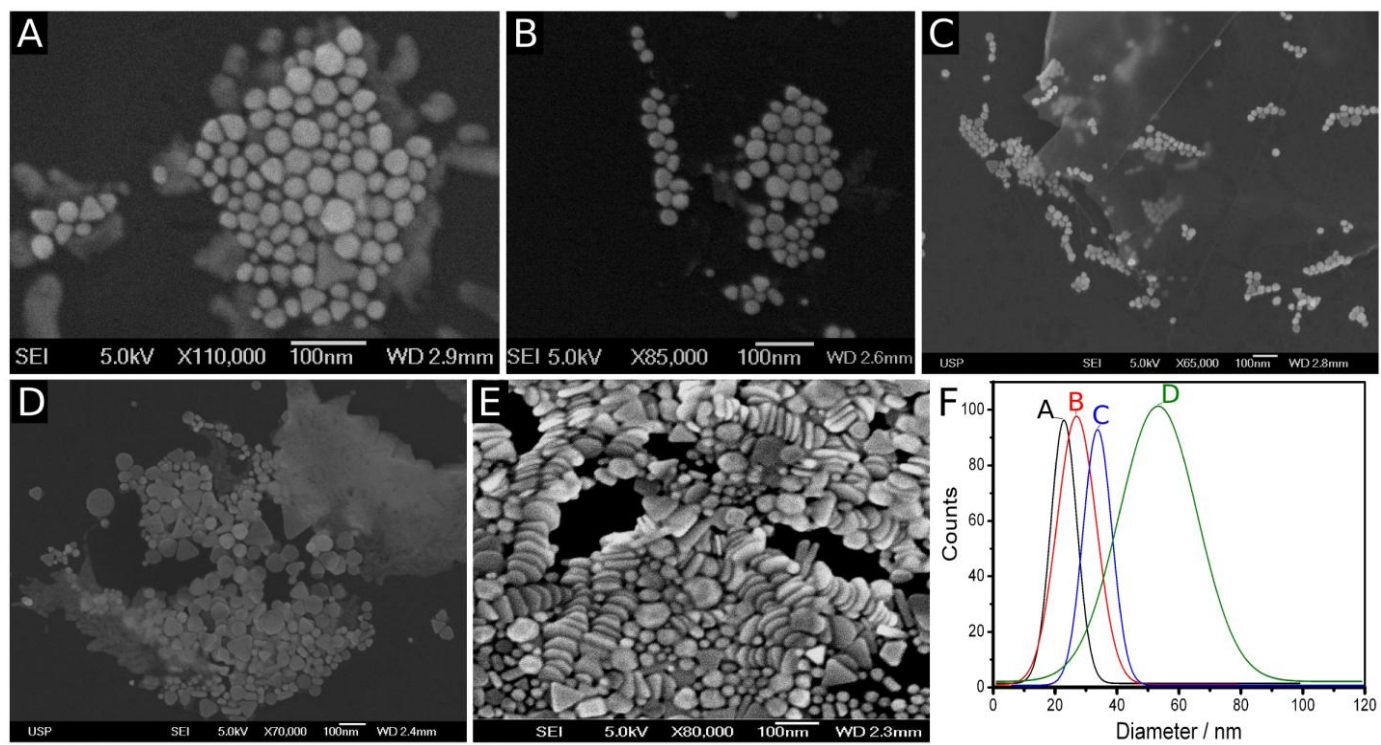

**Figure 11**. SEM images of: A) AuND-25; B) AuND-50; C) AuND-75; D) AuND-100; E) a different view of AuND-100; F) diameter distribution of the corresponding nanoparticles.

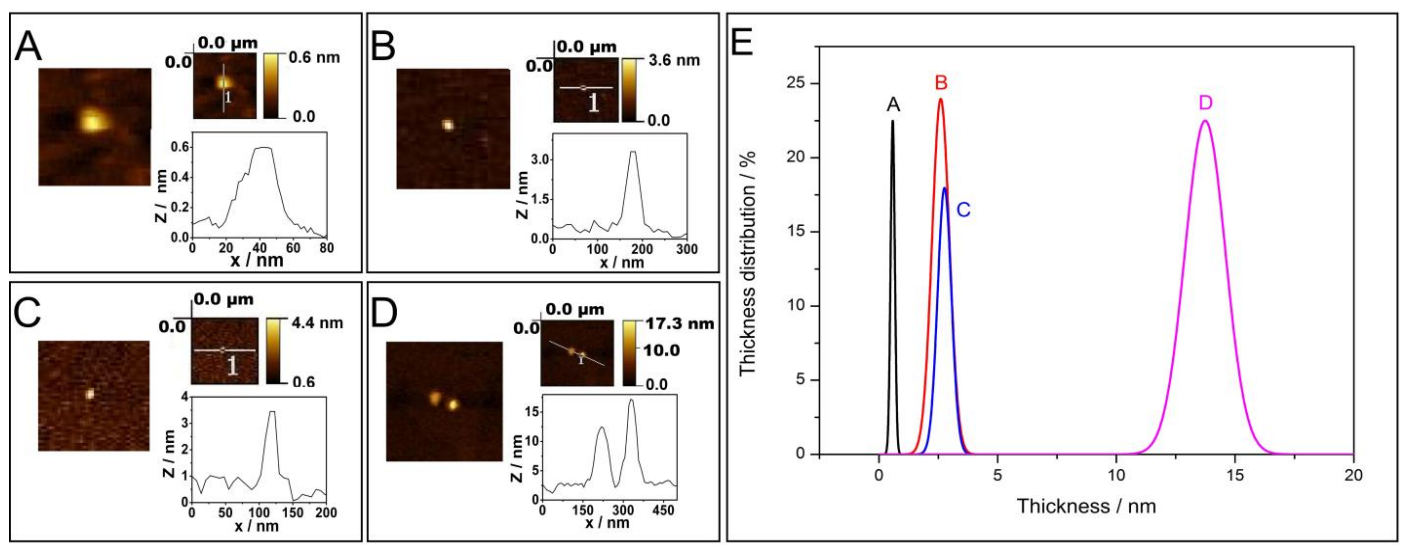

**Figure 12.** Representative topographic AFM images and the thickness (height, Z) of the A) AuND-25; B) AuND-50; C) AuND-75 and D) AuND-100 nanoparticles; E) thickness distribution of the corresponding AuND particles (averaged over 50 particles

## Conclusion

The formation of gold nanoparticles in micellar media is strongly dependent on the nature of the precursor complex and of the counter ions, while their geometries are related to the micelle shapes, which are not changed by the presence of the [AuCl$_4$]$^-$ complexes, accordingly to the SAXS measurements.

The precursor [AuCl$_4$]$^-$ complex is prone to substitution reactions in aqueous solution, but its preferential adsorption at the micellar interface prevents the occurrence of hydrolysis, since water concentration at the micellar interface is much smaller in relation to the bulk water concentration. However, chloride/bromide ligand exchange becomes important, influencing the redox processes and the formation of the gold nanoparticles. The poor coordinating properties of the Tf, Ms and TFA anions preclude the chloride/anion exchange, preserving the precursor complex identity, [AuCl$_4$]$^-$, in the micellar system. The Au reduction occurs at the micellar interface, and can be controlled by the counter ion nature. The [AuCl$_4$]$^-$ reduction with ascorbate proceeds at high kinetic rates, and the gold nanoparticles can be directly formed at the interfaces, with no need of adding nanoparticle seeds.

Bromide ions convert the [AuCl$_4$]$^-$ complex into [AuBr$_4$]$^-$, exhibiting a lower reduction potential, thus decreasing the electron transfer rates with the reducing agents (ascorbate). For this reason, no evidence of gold nanoparticles has been observed in DTA-Br solutions. In this case, the formation of gold nanoparticles requires the addition of seeds, in contrast to the systems involving non-coordinating counterions.

In summary, the counterion influences the kinetics of formation of gold nanoparticles by changing the gold complex redox potential, which controls the nucleation rate. As showed by our kinetic data the nucleation occur in one fast step flowed by one growth step. The counterion also control the micelle shapes, which in the case of non-coordinating ions control indirectly the nanoparticles shapes by promoting a directional grow of the nanoparticles tips in the case of DTA-Tf. In conclusion, the use of Br, Tf, Ms and TFA as DTA counter ion is a good example of kinetic control of the nanoparticle growth by controlling micellar and the gold complex properties by specific ion effects.

## Acknowledgements

The financial support from FAPESP (Projects 2011/00037-6, 13/08166-5 and 2014/50096-4), CNPq, Instituto Nacional de Fluidos Complexos (INCT-FCx), Núcleo de Apoio à Pesquisa de Fluidos Complexos, Núcleo de Apoio à Pesquisa em Nanotecnologia e Nanociências (NAP-NN) and PETROBRAS are gratefully acknowledged.

## Notes

The authors declare no competing financial interest. The synthesis of gold nanoparticles formation reported here is patented under the number BR 10 2014 013913-3.